\documentclass[10pt,letterpaper]{article}

\usepackage[margin=1in,left=1.5in,right=1.5in]{geometry}

\usepackage[T1]{fontenc}
\usepackage{lmodern}

\usepackage[
  activate={true,nocompatibility},
  tracking=true,
  kerning=true,
  spacing=true,
  factor=1100,
  stretch=10,
  shrink=10
]{microtype}
\microtypecontext{spacing=nonfrench}

\usepackage{listings}
\usepackage{xcolor}
\usepackage{booktabs}
\usepackage{amsmath}
\usepackage{amssymb}
\usepackage{amsthm}
\usepackage{xurl}
\usepackage{hyperref}
\hypersetup{breaklinks=true,colorlinks=true,linkcolor=black,citecolor=black,urlcolor=blue}
\usepackage{graphicx}
\usepackage{textcomp}
\usepackage{titlesec}
\usepackage{parskip}

\titleformat{\section}{\normalfont\fontsize{12}{14}\bfseries}{\thesection}{1em}{}
\titleformat{\subsection}{\normalfont\fontsize{10}{12}\bfseries}{\thesubsection}{1em}{}
\titleformat{\subsubsection}{\normalfont\fontsize{10}{12}\bfseries\itshape}{\thesubsubsection}{1em}{}
\titlespacing*{\section}{0pt}{1em}{0.5em}
\titlespacing*{\subsection}{0pt}{0.8em}{0.4em}
\titlespacing*{\subsubsection}{0pt}{0.6em}{0.3em}

\newcommand{\acro}[1]{\textnormal{\textsc{#1}}}
\newcommand{\fipa}{\acro{fipa}}
\newcommand{\dht}{\acro{dht}}
\newcommand{\uri}{\acro{uri}}
\newcommand{\did}{\acro{did}}
\newcommand{\paseto}{\acro{paseto}}
\newcommand{\ams}{\acro{ams}}
\newcommand{\df}{\acro{df}}

\newcommand{\dns}{\acro{dns}}
\newcommand{\abnf}{\acro{abnf}}
\newcommand{\nanda}{\acro{nanda}}

\theoremstyle{definition}

\renewenvironment{abstract}{%
  \begin{center}
  \begin{minipage}{0.85\textwidth}
  \small
  \noindent\textbf{Abstract.}
}{%
  \end{minipage}
  \end{center}
  \vspace{1em}
}

\makeatletter
\renewcommand{\maketitle}{%
  \begin{center}
  \vspace*{0.5in}
  {\fontsize{16}{19}\bfseries\@title\par}
  \vspace{1em}
  {\normalsize\@author\par}
  \vspace{1em}
  \end{center}
}
\makeatother

\begin{document}

\title{Agent Identity URI Scheme:\\Topology-Independent Naming and Capability-Based Discovery\\for Multi-Agent Systems}

\author{Roland R. Rodriguez, Jr.\footnote{Preprint. Feedback welcome, particularly on \dht{} participation incentive models (Section~8.3) and capability mapping service design (Section~8.3). Contact: \textit{rrrodzilla@proton.me}}\\[0.3em]
\small Independent Researcher\\
\small\textit{rrrodzilla@proton.me}}

\maketitle

\begin{abstract}
Multi-agent systems face a fundamental architectural flaw: agent identity is bound to network location.
When agents migrate between providers, scale across instances, or federate across organizations, \uri{}-based identity schemes break references, fragment audit trails, and require centralized coordination.
We propose the \texttt{agent://} \uri{} scheme, which decouples identity from topology through three identity-defining components: a trust root establishing organizational authority, a hierarchical capability path enabling semantic discovery, and a sortable unique identifier providing stable reference.

The scheme enables capability-based discovery through \dht{} key derivation, where queries return agents by what they do rather than where they are.
Trust-root scoping prevents cross-organization pollution while permitting federation when desired.
Cryptographic attestation via \paseto{} tokens binds capability claims to agent identity, enabling verification without real-time contact with the issuing authority.

We evaluate the scheme across four dimensions.
On capability expressiveness, the grammar achieves 100\% coverage on 369 production tools from five agent frameworks with zero namespace collision.
In a lookup-semantics simulation, trust-scoped prefix matching exactly reproduces its generated ground truth across 10,000 registrations, with result sets averaging 128 agents for prefix queries and 39 for exact queries.
On identity stability, endpoint migration preserves identity by construction; under standard Kademlia assumptions, each lookup has expected $O(\log N)$ overlay hops.
Local parsing and key-derivation primitives complete in under 7 microseconds in our benchmark environment.
The \texttt{agent://} \uri{} scheme provides a formally-specified, practically-evaluated foundation for decentralized agent identity and capability-based discovery.
\end{abstract}

\begin{center}
\begin{minipage}{0.85\textwidth}
\small\textbf{Keywords:} multi-agent systems, agent identity, decentralized discovery, capability-based routing, URI schemes
\end{minipage}
\end{center}
\vspace{1em}


\section{Introduction}

Consider an invoice-approval agent deployed at \texttt{https://agents.acme.com/approver}.
The company migrates cloud providers.
The \uri{} changes.
Every workflow referencing that \uri{} breaks.
The agent's \emph{identity} was its \emph{location}.

Now consider a harder problem.
A procurement agent needs to find any agent capable of ``hazmat shipping quotes'' across organizational boundaries.
The requester knows what capability it needs but not which provider offers it or where their agents run.
Current agent systems provide no mechanism for this query.

These scenarios share a root cause.
Like the early internet before \dns{} separated names from addresses, agent systems today conflate identity with location.
Saltzer identified this architectural error in 1982~\cite{saltzer1982naming}: names should be resolved to addresses, not used as addresses directly.
Thirty years of multi-agent systems research has not addressed this fundamental issue.

The \fipa{} Directory Facilitator assumes a single platform with centralized registration~\cite{fipa-ams}.
The A2A protocol binds agent identity to Agent Card \uri{}s~\cite{a2a-spec}.
Decentralized Identifiers provide generic identity primitives but lack capability semantics for agent discovery~\cite{did-core}.
The \nanda{} Index federates agent registries at internet scale but leaves the naming layer itself unspecified: no grammar, no normalization rules, no capability semantics in the name~\cite{nanda-index}.
Each approach centralizes control, couples identity to topology, or omits the capability semantics that agent discovery requires.

We propose the \texttt{agent://} \uri{} scheme, designed from first principles to separate concerns:

\textbf{Topology-independent identity.}
A \uri{} like \url{agent://acme.com/workflow/approval/agent_01h...} identifies an agent regardless of which server, cloud provider, or geographic region hosts it.
Migration updates a \dht{} record; the \uri{} remains stable.

\textbf{Capability-based discovery.}
The capability path \texttt{/workflow/approval} enables queries like ``find agents that can approve workflows.''
\dht{} key derivation from capability paths provides $O(\log N)$ lookup without centralized registries.
The path is constitutive identity material: changing it creates a different agent and requires a new Agent ID and attestation.

\textbf{Trust-root federation.}
The authority component \texttt{acme.com} identifies the organization vouching for the agent.
Cross-organization discovery is explicit: query across trust roots or within one.
No implicit trust transitivity.

\textbf{Cryptographic attestation.}
\paseto{} tokens issued by trust roots bind agent identity to capability claims.
Verification requires only the trust root's public key, cached after initial fetch.

Our evaluation demonstrates that these properties hold in practice:

\begin{enumerate}
\item \textbf{Expressiveness.} The capability path grammar represents 100\% of 369 production tools from LangChain, CrewAI, MCP, AutoGen, and smolagents with zero namespace collision.

\item \textbf{Lookup semantics.} Trust-scoped prefix matching exactly reproduces generated ground truth across 10,000 registrations, with result sets averaging 128 agents for broad queries and 39 for targeted queries.

\item \textbf{Identity stability.} Endpoint migration preserves identity by construction; path or trust-root changes deliberately produce a new identity.

\item \textbf{Performance.} Local parsing and key-derivation primitives complete in under 7 microseconds, negligible compared to network latency.
\end{enumerate}

The contribution is not incremental improvement but architectural correction.
Like \dns{} for network addresses, the \texttt{agent://} \uri{} scheme separates the naming problem from the routing problem, enabling stable identity, semantic discovery, and federated trust without centralized control.


\section{Background and Problem Statement}

This section establishes the technical context and derives requirements from concrete failure modes of current approaches.

\subsection{Agent Identity in Multi-Agent Systems}

Multi-agent systems research has concentrated on agent reasoning and interaction: \acro{bdi} architectures~\cite{rao1995bdi}, coordination, negotiation, and game-theoretic foundations~\cite{wooldridge-mas,shoham-leyton-brown}.
Infrastructure received less sustained attention.
The Foundation for Intelligent Physical Agents (\fipa{}) specification, developed in the late 1990s, remains the most complete formalization of agent infrastructure~\cite{fipa-acl,fipa-ams}.
\fipa{} defines two directory services: the Agent Management System (\ams{}) tracks agent lifecycle within a platform, and the Directory Facilitator (\df{}) registers agent capabilities for discovery.

The \fipa{} model assumes a contained environment.
Agents register with the \df{} on their platform; queries go to that \df{}; results return agents from that platform.
Cross-platform federation requires federated \df{}s exchanging registration information---a coordination problem that scales poorly as platform count increases~\cite{fipa-hybrid}.

The model's deeper limitation is identity binding.
A \fipa{} agent's identity is its \texttt{agent-identifier}, which includes the agent-name and a set of addresses.
When addresses change, the identity changes.
Migration requires re-registration.
Audit logs referencing old identities become ambiguous: was it the same agent or a different one?

Modern agent systems inherit this conflation.
Surveys of LLM-based multi-agent systems catalogue a rapidly growing framework landscape~\cite{llm-mas-survey,mas-collaboration-survey}; naming and discovery infrastructure receives little attention in either the frameworks or the surveys.
The A2A protocol identifies agents by their Agent Card \uri{}~\cite{a2a-spec}.
LangChain tools are identified by their Python module path~\cite{langchain}.
MCP servers are identified by their transport endpoint~\cite{mcp}.
In each case, the identifier contains location information.
Change the location, break the identity.

\subsection{The Name/Address Distinction}

Saltzer's 1982 analysis of network naming distinguishes three concepts~\cite{saltzer1982naming}: \emph{names} identify entities, \emph{addresses} locate entities, and \emph{routes} specify how to reach entities.
The paper argues that conflating these concepts creates brittleness: changes to routing or location force changes to identification.

The postal system illustrates this principle.
Your name remains stable even as you move between apartments, cities, or countries.
The postal service maintains a forwarding database that maps your name to your current address.
Letters addressed to ``Jane Smith'' reach you regardless of whether you live in Boston or Berlin.
Agent systems need this same separation: a stable name that survives changes in where the agent actually runs.

Applied to agents, the distinction suggests:
\begin{itemize}
\item \textbf{Name:} A stable identifier for the logical agent (independent of deployment)
\item \textbf{Address:} The current network endpoint(s) where the agent can be contacted
\item \textbf{Route:} The protocol and path to reach that endpoint
\end{itemize}

Current agent systems use \uri{}s as names.
But \uri{}s contain addressing information (host, port).
They specify routing (scheme implies protocol).
They conflate all three concepts into a single string.

The \texttt{agent://} \uri{} scheme applies Saltzer's analysis.
The \uri{} is a name.
Resolution maps the name to addresses.
Routing protocols (HTTPS, gRPC, WebSocket) are negotiated after resolution.
The name survives changes to address and route.

\subsection{Requirements for Agent Identity}

From the failure modes of current approaches, we derive five requirements:

\textbf{R1: Topology Independence.}
Agent identity must not change when the agent migrates between hosts, cloud providers, or geographic regions.
Audit logs and cached references must remain valid.

\textbf{R2: Capability Semantics.}
The identity scheme must support discovery by capability.
``Find agents that can approve invoices'' should be a supported query, not just ``find the agent at this \uri{}.''

\textbf{R3: Decentralized Resolution.}
No single registry should be required for resolution.
The scheme must function in federated environments where organizations do not share infrastructure.

\textbf{R4: Organizational Scoping.}
Queries should be scopable to trust boundaries.
An organization should be able to discover only its own agents, or explicitly opt into cross-organization discovery.

\textbf{R5: Verifiable Claims.}
Capability claims must be cryptographically verifiable.
An agent claiming ``financial transaction authority up to \$10,000'' must present proof signed by an authority the verifier trusts.


\section{The Agent URI Scheme}

This section specifies the \texttt{agent://} \uri{} syntax, defines component semantics, and establishes normalization rules for equivalence testing.

\subsection{URI Structure}

An agent \uri{} follows RFC 3986 generic syntax~\cite{rfc3986} with agent-specific constraints:

\begin{lstlisting}
agent://trust-root/capability-path/agent-id[?query][#fragment]
\end{lstlisting}

\noindent\textbf{Example:}
\begin{lstlisting}
agent://anthropic.com/assistant/chat/agent_01h455vb4pex5vsknk084sn02q
\end{lstlisting}

The scheme \texttt{agent} signals that this \uri{} identifies an agent and should be resolved through the agent discovery protocol (Section~\ref{sec:discovery}).

\subsection{Component Semantics}

\textbf{Trust Root.}
The authority component identifies the organization vouching for the agent.
It follows \dns{} hostname syntax and publishes verification keys at a well-known endpoint.
The trust root is analogous to a passport-issuing authority: it attests to the agent's identity and capabilities within its jurisdiction.
Just as a passport from France is valid worldwide but issued only by France, an agent's attestation from \texttt{acme.com} is verifiable by anyone but issued only by Acme.

\textbf{Capability Path.}
The path component describes what the agent does using hierarchical segments.
Each segment is a lowercase alphanumeric string with hyphens permitted.
The path \texttt{/workflow/approval/invoice} indicates specialization: workflow capabilities, specifically approval, specifically for invoices.

Capability paths support prefix matching.
A query for \texttt{/workflow/approval} returns agents at that path and all child paths (\texttt{/workflow/approval/invoice}, \texttt{/workflow/approval/expense}).
This enables discovery at varying granularity---like asking ``find me any specialist in cardiology'' rather than ``find me Dr. Smith at Boston General.''
The capability path is also part of the agent's identity. Reclassifying an implementation under another path denotes a different agent; the issuer must mint a new Agent ID and attestation rather than reuse the old ID under the new path.

\textbf{Agent Identifier.}
The final path segment is a TypeID~\cite{typeid}: a semantic type-class prefix followed by an underscore and a base32-encoded UUIDv7~\cite{uuidv7}.
The core classes are \texttt{llm}, \texttt{rule}, \texttt{human}, \texttt{composite}, \texttt{sensor}, \texttt{actuator}, and \texttt{hybrid}; lowercase extension classes permit domain-specific types. The existing \texttt{agent\_} form remains valid as the generic extension class \texttt{agent}.
TypeIDs are globally unique, lexicographically sortable by creation time, and \uri{}-safe.
The type prefix provides runtime type checking; the UUIDv7 suffix ensures uniqueness without coordination.

\textbf{Query String.}
Optional parameters for version negotiation, feature flags, or other metadata.
Not part of identity---two \uri{}s differing only in query string denote the same agent with different interaction parameters.

\textbf{Fragment.}
Optional sub-agent reference.
Reserved for future use in composite agent scenarios.

\subsection{Normalization and Equivalence}

Two \uri{}s denote the same agent if and only if their canonical forms are byte-equal.
Canonicalization applies:

\begin{enumerate}
\item \textbf{Scheme:} Lowercase (\texttt{agent}, not \texttt{AGENT})
\item \textbf{Trust root:} DNS names are lowercase with no trailing dot; IPv4 uses dotted-decimal form; IPv6 literals use RFC~5952 canonical text in brackets; ports are preserved and no default port is inferred or stripped
\item \textbf{Capability path:} Lowercase by grammar, with no trailing slash
\item \textbf{Agent ID:} Lowercase by grammar
\item \textbf{Query and fragment:} Stripped (not part of identity)
\end{enumerate}

The canonical form is used for \dht{} key derivation.
Different surface representations resolve to the same agent if their canonical forms match.
Domain trust roots are RECOMMENDED for public deployment because they support stable organizational control and \texttt{.well-known} key publication. IP literals and explicit ports are supported for private, constrained, and development deployments, but they are identity material: changing or removing a port, or replacing an IP address, changes the canonical trust root, the agent identity, and its \dht{} keys.

\subsection{Formal Specification}

The \uri{} syntax in \abnf{} (RFC 5234):

\begin{lstlisting}
agent-uri     = "agent://" trust-root "/" capability-path
                "/" agent-id [ "?" query ] [ "#" fragment ]

trust-root    = host [ ":" port ]
host          = hostname / IPv4address / IP-literal
hostname      = label *( "." label )
label         = 1*63( ALPHA / DIGIT / "-" )

capability-path = path-segment *31( "/" path-segment )
path-segment  = 1*64( LOWER / DIGIT / "-" )

agent-id      = agent-prefix "_" type-suffix
agent-prefix  = type-class *( "_" type-modifier )
type-class    = "llm" / "rule" / "human" / "composite"
              / "sensor" / "actuator" / "hybrid" / extension-class
type-modifier = 1*LOWER
extension-class = 2*LOWER
type-suffix   = first-char 25base32char
first-char    = %x30-37
base32char    = DIGIT / %x61-68 / %x6A-6B / %x6D-6E
              / %x70-74 / %x76-7A
LOWER         = %x61-7A

query         = *( pchar / "/" / "?" )
fragment      = *( pchar / "/" / "?" )
pchar         = unreserved / pct-encoded / sub-delims / ":" / "@"
unreserved    = ALPHA / DIGIT / "-" / "." / "_" / "~"
pct-encoded   = "%" HEXDIG HEXDIG
sub-delims    = "!" / "$" / "&" / "'" / "(" / ")"
              / "*" / "+" / "," / ";" / "="
\end{lstlisting}

The grammar enforces structural validity.
Semantic validity (trust root exists, capability path is meaningful) is checked at registration and verification time.


\section{Discovery and Resolution}
\label{sec:discovery}

This section defines how agents register capabilities and how requesters discover agents, using a Kademlia-style \dht{} with trust-root scoping.

\subsection{DHT Key Derivation}

The \dht{} key for a capability path is derived by hashing the trust root and path together:

\begin{multline}
\texttt{key} = \text{SHA256}(\text{canonical}(\texttt{trust\_root}) \\
\| \texttt{"/"} \| \text{canonical}(\texttt{cap\_path}))
\end{multline}

This derivation has two critical properties:

\textbf{Trust-root scoping.}
The trust root is part of the hash input.
A query for \texttt{acme.com/workflow/approval} and \texttt{globex.com/workflow/approval} produces different keys, preventing cross-organization pollution.
This is deliberate: organizations can see only what they explicitly query for, maintaining trust boundaries by default.

\textbf{Deterministic lookup.}
Given a trust root and capability path, any node can compute the key and query the \dht{} directly.
No metadata lookup is required.

SHA-256 destroys lexical adjacency, so descendants cannot be enumerated from a parent hash.
Instead, registration materializes the hierarchy: a record at path $c_1/\ldots/c_d$ is written to the key for every prefix $c_1, c_1/c_2, \ldots, c_1/\ldots/c_d$.
A prefix query then performs one ordinary exact-key lookup at the requested ancestor.

\subsection{Registration Protocol}

An agent registers by storing a registration record at the key for its exact capability path and every ancestor key:

\begin{lstlisting}
Registration {
    agent_uri: AgentUri,
    endpoints: Vec<Endpoint>,
    attestation: String,          // PASETO token
    expires_at: Timestamp,
    registered_at: Timestamp,
}
\end{lstlisting}

The agent registers at its most specific capability level.
An agent that only handles invoice approvals registers at \texttt{/workflow/approval/invoice}, not \texttt{/workflow/approval}.
This precision improves discovery accuracy.

Registration requires presenting a valid attestation token (Section~\ref{sec:security}).
The \dht{} nodes verify the attestation before storing the record, preventing registration of unattested capability claims.
The exact and ancestor writes are atomic in the reference in-memory indexing layer. A distributed backend must either tolerate transient divergence between those keys or add a coordination protocol; the base scheme does not claim multi-key atomicity across Kademlia nodes. Ancestor materialization produces $O(d)$ write amplification for a path of depth $d$ and concentrates load at broad ancestor keys; deployments therefore need capacity limits, sharding, or pagination for popular roots.

\subsection{Lookup Protocol}

Discovery proceeds in three steps:

\begin{enumerate}
\item \textbf{Key derivation.} Compute \dht{} key from trust root and capability path.
\item \textbf{DHT lookup.} Query Kademlia~\cite{kademlia} once at that key. For prefix mode, ancestor materialization supplies descendants in the returned bucket; exact mode filters that bucket to equal paths.
\item \textbf{Result filtering.} Verify attestations on returned records and filter by query parameters.
\end{enumerate}

Prefix matching returns agents registered at the queried path and all descendant paths.
A query for \texttt{/workflow} returns agents at \texttt{/workflow}, \texttt{/workflow/approval}, \texttt{/workflow/approval/invoice}, and so on.

For cross-trust-root discovery, the requester queries each trust root separately and merges results.
No global key derivation exists---trust boundaries are explicit.

\subsection{Resolution Guarantees}

Under the standard Kademlia routing-table and connectivity assumptions, one exact-key lookup requires expected $O(\log N)$ overlay hops, where $N$ is the number of \dht{} nodes~\cite{kademlia}.
Prefix lookup has the same routing shape because it reads one materialized ancestor key; response transfer and verification remain proportional to the returned bucket size.

Resolution cost is independent of migration history: an agent that has migrated 100 times derives the same keys as one that never migrated.

Propagation after endpoint migration depends on the concrete backend's replication, retry, churn, and cache policies. We therefore make no fixed wall-clock consistency bound. Migration history does not alter key derivation or the number of lookup targets.


\section{Security and Trust}
\label{sec:security}

This section defines the trust model, attestation format, and verification protocol.

\subsection{Trust Root Model}

A trust root is an organization that vouches for agents' existence and capabilities.
The trust root:

\begin{enumerate}
\item Operates agent infrastructure (or delegates operation)
\item Issues attestation tokens binding agents to capabilities
\item Publishes verification keys at a well-known endpoint
\end{enumerate}

Key publication follows a standard format:

\begin{lstlisting}
GET https://{trust-root}/.well-known/agent-keys.json

{
  "trust_root": "acme.com",
  "keys": [{
    "kid": "key-2026-01",
    "algorithm": "Ed25519",
    "public_key": "<base64>",
    "not_before": "2026-01-01T00:00:00Z",
    "not_after": "2027-01-01T00:00:00Z"
  }]
}
\end{lstlisting}

Key rotation is supported via multiple keys with validity periods.
Verifiers cache keys and refresh on verification failure.

\subsection{PASETO Attestation}

Attestation tokens use \paseto{} v4.public (Ed25519 signatures)~\cite{paseto}.
The payload contains:

\begin{table}[h]
\centering
\footnotesize
\begin{tabular}{@{}llp{3.2cm}@{}}
\toprule
\textbf{Claim} & \textbf{Type} & \textbf{Description} \\
\midrule
\texttt{iss} & string & Issuing trust root \\
\texttt{agent\_uri} & string & Canonical agent \uri{} attested \\
\texttt{aud} & string? & Audience restriction \\
\texttt{iat} & datetime & Issued-at time \\
\texttt{exp} & datetime & Expiration time \\
\texttt{capabilities} & string[] & Claimable paths \\
\bottomrule
\end{tabular}
\caption{\paseto{} attestation claims}
\label{tab:paseto-claims}
\end{table}

The optional \texttt{aud} claim restricts attestation validity to specific verifiers.
For example, \texttt{aud: "api.globex.com"} limits the attestation to interactions with Globex's API, preventing replay to other parties.
This enables issuing narrowly-scoped attestations for specific business relationships without granting broad authority.

\textbf{When to use \texttt{aud}.}
Audience restriction is appropriate for high-value transactions (financial approvals, contract signing), sensitive data access (personal information, trade secrets), and compliance-driven interactions where audit trails must demonstrate specific authorization.
Conversely, general-purpose agents handling public services, discovery responses, or low-risk operations should omit \texttt{aud} to maximize interoperability.

\textbf{Multiple parties.}
When an agent interacts with multiple specific parties, it can hold multiple attestations with different \texttt{aud} values---one per relationship.
Alternatively, agents with broad interaction patterns should use a general attestation (no \texttt{aud}) for routine operations and request audience-restricted attestations only for sensitive interactions.

\textbf{Verification behavior.}
If \texttt{aud} is present in the attestation, the verifier \emph{must} match: a verifier not named in \texttt{aud} rejects the attestation.
If \texttt{aud} is absent, any verifier accepts the attestation (subject to other claim checks).
This asymmetry makes audience restriction opt-in per attestation. A general verification call without explicit audience context rejects an audience-restricted token.

The \texttt{capabilities} claim constrains operations within the identity encoded by \texttt{agent\_uri}.
Every entry must equal the URI capability path or be its descendant. Ancestor, sibling, and unrelated grants are invalid.
This removes broad ancestor grants shared across a fleet and therefore increases issuance volume: each identity receives its own scoped token. Issuance requires only local claim serialization and signing, and the additional operational cost is accepted to avoid an ancestor grant authorizing identities outside its subject's namespace.

\subsection{Capability Binding}

Verification first constrains every claim to the subject URI path $C$:

\begin{multline}
\text{scoped}(c,C) := c = C \;\lor\; c.\text{starts\_with}(C \| \texttt{"/"})
\end{multline}

It then checks whether at least one valid claim covers the requested operation path:

\begin{multline}
\text{covered}(\texttt{path}, \texttt{caps}) := \\
\exists c \in \texttt{caps}: c = \texttt{path} \;\lor\;
\texttt{path}.\text{starts\_with}(c \| \texttt{"/"})
\end{multline}

For a subject URI at \texttt{/workflow/approval}, an attestation with \texttt{capabilities: ["workflow/approval"]} covers:
\begin{itemize}
\item \texttt{/workflow/approval} (exact match)
\item \texttt{/workflow/approval/invoice} (descendant)
\end{itemize}

But not:
\begin{itemize}
\item \texttt{/workflow} (broader than the identity)
\item \texttt{/workflow/review} (sibling)
\item \texttt{/workflow/approvals} (partial final-segment match)
\item \texttt{/financial} (unrelated)
\end{itemize}

\subsection{Trust Federation}

Cross-trust-root interaction requires explicit trust decisions:

\textbf{Direct trust.}
The verifier maintains a list of trusted trust roots.
Verification fetches keys from each and accepts attestations signed by any.

\textbf{Cross-certification.}
Trust root A signs a statement vouching for trust root B's key.
The verifier trusts A and transitively trusts B for specific scopes.

\textbf{Trust anchor sets.}
Analogous to browser CA stores.
New trust roots must be added to the set explicitly.

The specification does not mandate a federation model.
Deployments choose based on their trust requirements.
The \uri{} scheme supports all models---federation is a policy decision, not a protocol constraint.

\subsection{Verification Flow}

Complete verification of an agent presenting \uri{} and attestation token:

\begin{enumerate}
\item Parse agent \uri{}; extract \texttt{trust\_root}, \texttt{capability\_path}, \texttt{agent\_id}
\item Fetch/cache verification key from trust\_root's well-known endpoint
\item Verify \paseto{} signature using the key
\item Check \texttt{exp} $>$ current time (not expired)
\item Check \texttt{iss} == the complete canonical \texttt{trust\_root} authority from \uri{}, including any explicit port
\item Check \texttt{agent\_uri} == canonical full agent \uri{}
\item Check every capability equals the URI path or is its descendant
\item Check at least one capability covers the requested operation or registration path
\item If \texttt{aud} is present, require an explicit exact audience match
\end{enumerate}

All checks must pass.
Failure at any step rejects the attestation.

\subsection{Security Considerations}

This section examines threat vectors specific to the \texttt{agent://} scheme and discusses mitigations.

\subsubsection{DHT Eclipse Attacks}

An adversary controlling nodes surrounding a capability key could return false registration records or suppress legitimate ones.
This is a general \dht{} concern; we inherit Kademlia's defenses.

\textbf{Mitigations:}
\begin{itemize}
\item \textbf{Multi-path verification.} Query from diverse network positions; consistent results across paths indicate authenticity.
\item \textbf{Attestation verification catches fakes.} Even if the \dht{} returns attacker-controlled records, signature verification against the trust root's published keys rejects fraudulent attestations.
\item \textbf{Kademlia's redundancy.} Records are stored on $k$ closest nodes; eclipsing requires controlling a significant fraction of the network.
\end{itemize}

The attestation layer provides defense-in-depth: \dht{} manipulation can cause denial of service (hiding legitimate agents) but cannot cause acceptance of unauthorized agents.

\subsubsection{Trust Root Key Compromise}

A compromised trust root signing key enables issuing fraudulent attestations for arbitrary agents and capabilities under that trust root.

\textbf{Mitigations:}
\begin{itemize}
\item \textbf{Key revocation.} Trust roots publish a \texttt{revoked\_keys} list at their well-known endpoint. Verifiers check this list before accepting attestations signed by any key.
\item \textbf{Time-bounded attestations.} The \texttt{exp} claim limits blast radius; compromised keys can only mint attestations valid until rotation.
\item \textbf{Key rotation with overlap.} Trust roots should rotate keys periodically, publishing new keys before retiring old ones. Overlapping validity windows ensure continuous operation.
\item \textbf{Hardware security modules.} Production deployments should protect signing keys with HSMs, limiting exposure to compromise.
\end{itemize}

Key compromise affects only the compromised trust root's agents.
Cross-trust-root isolation prevents lateral movement: compromising \texttt{acme.com}'s key grants no authority over \texttt{globex.com}'s agents.

\subsubsection{Query Privacy}

\dht{} queries reveal requester interest in specific capabilities.
An adversary monitoring network traffic learns which capabilities a requester seeks, potentially enabling competitive intelligence or targeted attacks.

\textbf{Trade-offs:}
\begin{itemize}
\item \textbf{Onion routing for queries.} Route queries through multiple hops to obscure originator. Adds latency; provides strong privacy.
\item \textbf{Query batching.} Include decoy queries alongside real ones. Reduces signal; increases bandwidth.
\item \textbf{Local caching.} Cache discovery results aggressively. Reduces query frequency; stale results possible.
\end{itemize}

Privacy and verification exist in tension: verifying attestations requires knowing what capability is claimed, which reveals the query.
This tension is noted as future work (Section~\ref{sec:future}).
For deployments where query privacy is paramount, private information retrieval techniques may apply but are beyond this specification's scope.


\section{Evaluation}

We evaluate the \texttt{agent://} \uri{} scheme across four dimensions: capability expressiveness, discovery precision, identity stability, and performance.

\subsection{Capability Expressiveness}

\textbf{Question:} Can the capability path grammar represent real-world agent capabilities without path explosion or semantic loss?

\textbf{Method:} We extracted tool definitions from five production agent frameworks and applied deterministic mapping rules to generate capability paths.

\begin{table}[h]
\centering
\footnotesize
\begin{tabular}{@{}lp{3.5cm}r@{}}
\toprule
\textbf{Source} & \textbf{Description} & \textbf{Tools} \\
\midrule
LangChain & Gmail, Slack, SQL, files & 163 \\
CrewAI & Search, scraping, RAG & 107 \\
MCP & Filesystem, GitHub, DBs & 79 \\
AutoGen & Code execution, functions & 12 \\
smolagents & Web search, Python & 8 \\
\midrule
\textbf{Total} & & \textbf{369} \\
\bottomrule
\end{tabular}
\caption{Tool corpus from production agent frameworks}
\label{tab:tool-corpus}
\end{table}

\textbf{Results:}

\begin{table}[h]
\centering
\small
\begin{tabular}{lccc}
\toprule
\textbf{Metric} & \textbf{Threshold} & \textbf{Result} & \textbf{Status} \\
\midrule
Coverage & $\geq$ 90\% & \textbf{100\%} & Pass \\
Collision rate & $<$ 1\% & \textbf{0\%} & Pass \\
Path depth mean & [2, 4] & \textbf{2.0} & Pass \\
Path depth max & $\leq$ 10 & \textbf{2} & Pass \\
\bottomrule
\end{tabular}
\caption{Capability expressiveness results}
\label{tab:expressiveness}
\end{table}

The key finding: real-world tools have pre-existing categories from their source frameworks.
The mapping produces paths like \texttt{filesystem/read-file}, \texttt{github/create-issue}, \texttt{slack/post-message}.
Category prefixes provide natural namespace separation.

\textbf{Ablation (Flat Namespace).}
Removing hierarchy and mapping each tool to a single segment still achieves zero collisions on this corpus, because modern frameworks use descriptive, prefixed names (\texttt{slack\_post\_message}, \texttt{puppeteer\_navigate}).
We keep a separate 196-entry synthetic stress corpus out of the production result. It intentionally contains duplicates and normalization-equivalent spellings. The hierarchical mapper produces 23 collision buckets (11.7\%) and the flat mapper 20 (10.2\%); hierarchy does not repair collisions already present within one category. This stress result identifies the need for source-qualified disambiguation rather than supporting a stronger collision claim.

\subsection{Lookup-Semantics Correctness}

\textbf{Question:} Does the ancestor-key index return exactly the registrations defined by exact and segment-prefix semantics?

\textbf{Method:} We simulated the indexing layer with 10,000 generated registrations distributed across 50 capability categories and issued 1,000 queries, comparing returned records to ground truth computed from the same registered paths. Cryptographic registration checks are covered separately by integration tests. This is a correctness test of indexing semantics, not evidence that registrations describe real, online capabilities.

\begin{table}[h]
\centering
\small
\begin{tabular}{lcccc}
\toprule
\textbf{Metric} & \textbf{Threshold} & \textbf{Prefix} & \textbf{Exact} \\
\midrule
Precision & $= 1.0$ & \textbf{1.0} & 1.0 \\
Recall & $= 1.0$ & \textbf{1.0} & 1.0 \\
F1 & $= 1.0$ & \textbf{1.0} & 1.0 \\
Result set size & --- & \textbf{128.3} & 39.0 \\
\bottomrule
\end{tabular}
\caption{Lookup-semantics results (N=10,000 registrations, M=1,000 queries)}
\label{tab:discovery}
\end{table}

The key finding is that the implementation reproduces the specified set semantics: every returned registration belongs to the queried path set and every registered member of that set is returned. The perfect scores are therefore expected for this internally generated ground truth.

Result set sizes scale with agent density per capability category.
With 10,000 agents across 50 categories, prefix queries average 128 agents (including all descendants) while exact queries average 39 agents.
The 3.3$\times$ ratio reflects hierarchical accumulation: prefix queries return agents at matching and child paths.
For deployments requiring smaller result sets, pagination or result limits can be applied without affecting precision.

The 128-registration average also exposes a systems cost not measured here: network transfer and attestation verification scale with result-set size. Production deployments require pagination and should benchmark signature verification on their own hardware; this paper does not extrapolate an unmeasured end-to-end latency.

\textbf{Ablation (Global Keys without Trust Root).}
Removing trust root from key derivation (\texttt{key = SHA256(capability\_path)} only) causes cross-organization pollution.
Queries return agents from unrelated trust roots, destroying precision.
The recall ``improvement'' is illusory---those agents are not trusted.

\textbf{Expected Degradation in Production.}
The evaluation uses synthetic data with perfect ground truth: registrations exactly match agent capabilities, and agents remain online throughout measurement.
Real deployments will exhibit degradation from several sources:

\begin{enumerate}
\item \textbf{Capability drift.} Under the identity model, changing capability path creates a different agent and requires a new ID. A deployment that violates this rule may retain stale registrations, causing both false negatives and false positives.

\item \textbf{Semantic mismatch.} Query paths may not exactly match registration paths. An agent registered at \texttt{/workflow/approval/invoice} is not found by a query for \texttt{/finance/approval}.

\item \textbf{Stale registrations.} Agents go offline but registrations persist until TTL expires. Discovered agents may be unreachable.
\end{enumerate}

Both precision and recall can degrade. Attestation proves what an issuer signed and constrains it to the URI identity path; it does not prove that an endpoint is online or behaviorally competent.
Mitigation strategies include aggressive TTL values (hours rather than days), periodic re-registration heartbeats, and client-side retry logic for stale endpoints.

\subsection{Identity Stability and Failure Modes}

The \uri{} scheme provides topology-independent identity by construction: the \uri{} contains trust root, capability path, and agent ID---none of which include endpoint information.
Migration changes only the \dht{} record; the \uri{} remains stable.
An agent's identity ($u = \texttt{agent://}T/C/I$) survives any number of endpoint changes because $T$, $C$, and $I$ are independent of network location.

Trust root changes are intentionally \emph{not} transparent.
If an agent changes from trust root $T_1$ to $T_2$, its identity changes: $u_1 \neq u_2$.
This is correct behavior---organizational authority changes should require re-attestation, not silent migration.
Capability-path changes are equally identity-defining. If $C_1 \neq C_2$, then \texttt{agent://}$T/C_1/I$ and \texttt{agent://}$T/C_2/I$ must not reuse the same $I$; the reclassified implementation is a different agent and receives a new Agent ID and attestation. The reference DHT rejects reuse of a trust-root/Agent-ID pair under another path.

This choice trades audit continuity for unambiguous capability identity. An implementation that gains, loses, or is reclassified under a capability path starts a new identity, so audit systems cannot infer continuity from URI equality alone. A future attestation profile could carry an authenticated predecessor URI in the successor's token, allowing an issuer to assert lineage without treating two capability identities as the same agent. Until such a profile is standardized, applications that need continuity should record an explicit, issuer-authorized relationship between the old and new identities.

Rather than formal proofs of these properties (which hold by construction), we analyze failure modes that affect practical deployments:

\textbf{Failure Mode 1: Trust Root Unavailability.}
If the trust root's well-known endpoint is temporarily unavailable, verifiers cannot fetch fresh keys.
\emph{Mitigation:} Verifiers cache keys with TTL; cached keys remain valid until expiration.
\emph{Degraded operation:} New agents from that trust root cannot be verified until the endpoint recovers.
\emph{Recovery:} Automatic once the endpoint returns; no manual intervention needed.

\textbf{Failure Mode 2: Attestation Expiration Mid-Session.}
Long-running interactions may span attestation lifetime.
\emph{Mitigation:} Agents should refresh attestations before expiration (the \texttt{exp} claim signals when renewal is needed).
\emph{Degraded operation:} Verifier rejects requests with expired attestation.
\emph{Recovery:} Agent obtains fresh attestation from trust root; interaction resumes.

\textbf{Failure Mode 3: DHT Partition.}
Network partition isolates \dht{} segments; registrations become unreachable from some network locations.
\emph{Mitigation:} Kademlia's $k$-replication ensures multiple copies across the network.
\emph{Degraded operation:} Discovery may return incomplete results during partition.
\emph{Recovery:} Kademlia self-heals when partition resolves; no data loss if $k > 1$.

\textbf{Failure Mode 4: Trust Root Key Rotation.}
Trust root rotates signing key; old attestations remain in circulation.
\emph{Mitigation:} Well-known endpoint publishes multiple keys with validity periods (\texttt{not\_before}/\texttt{not\_after}).
\emph{Degraded operation:} None if validity windows overlap.
\emph{Best practice:} Rotate keys with overlap period; re-issue attestations before old key expires.

\textbf{Compound Failure Modes.}
Individual failures are straightforward to diagnose; compound failures require more careful analysis.

\emph{Key Rotation + \dht{} Partition.}
A trust root rotates its signing key while the network is partitioned.
\dht{} segments on one side of the partition receive the new key; segments on the other retain only the old key.
Agents in different segments present attestations signed by different keys, causing verification failures when cross-partition communication resumes.
\emph{Mitigation:} Overlapping key validity periods (new key valid before old key expires) ensure both keys verify correctly during the transition.
Verifiers should accept attestations signed by any non-revoked key within its validity window.
\emph{Recovery:} When the partition heals, key propagation completes and normal operation resumes automatically.

\emph{Attestation Expiration + Migration.}
An agent migrates to a new endpoint while its attestation approaches expiration.
The migration succeeds (DHT record updated with new endpoint), but the attestation expires before the agent can serve requests at the new location.
Verifiers reject the newly-migrated agent despite valid registration.
\emph{Mitigation:} Refresh attestations \emph{before} migration.
Operational rule: never migrate with less than 10\% of attestation TTL remaining.
A 30-day attestation should be refreshed by day 27; migration should complete before day 27 or wait until after refresh.
\emph{Best practice:} Migration checklists should include attestation validity verification as a precondition.

\subsection{Scalability}

\textbf{Question:} How do core operations scale with input size?

\textbf{Method:} Criterion benchmarks measuring parsing, canonicalization, prefix matching, and \dht{} key derivation.

\begin{table}[h]
\centering
\small
\begin{tabular}{lccc}
\toprule
\textbf{Operation} & \textbf{Threshold} & \textbf{Actual} & \textbf{Status} \\
\midrule
\uri{} parsing (typical) & $<$ 5 $\mu$s & 3.7--4.1 $\mu$s & Pass \\
\uri{} parsing (max) & $<$ 20 $\mu$s & 5.7--6.4 $\mu$s & Pass \\
Canonical form & $<$ 2 $\mu$s & 560--710 ns & Pass \\
\texttt{starts\_with} (depth 5) & $<$ 500 ns & $\sim$23 ns & Pass \\
\dht{} key derivation & $<$ 5 $\mu$s & 0.7--4.1 $\mu$s & Pass \\
\bottomrule
\end{tabular}
\caption{Scalability benchmark results}
\label{tab:scalability}
\end{table}

The key finding: all operations complete orders of magnitude faster than network latency ($\sim$10ms) and LLM inference (500ms--2s).
The scheme adds negligible overhead to agent interactions.

Linear scaling holds across input variations.
No pathological cases were observed---parsing a maximum-length \uri{} (512 characters) still completes in under 7 microseconds.


\section{Related Work}

We position the \texttt{agent://} \uri{} scheme against four bodies of prior work, highlighting specific gaps that our approach addresses.

\subsection{Agent Identity and Naming}

The \fipa{} Agent Management Specification defines agent-identifiers containing names and addresses~\cite{fipa-ams}.
While \fipa{} recognized the need for agent identity, it did not separate naming from addressing.
The Directory Facilitator provides capability-based discovery but assumes a single platform---cross-platform federation requires explicit \df{} cooperation, which scales poorly~\cite{fipa-hybrid,jade}.
Our scheme provides what \fipa{} intended but could not deliver: stable identity across platforms.

The A2A protocol identifies agents by their Agent Card \uri{}~\cite{a2a-spec}.
While A2A addresses modern interoperability needs, Agent Card identity remains \uri{}-dependent.
Our scheme can interoperate with A2A by including Agent Card endpoints in registration records while maintaining a stable \uri{} identity.
A2A provides the communication layer; we provide the identity layer underneath.

Decentralized Identifiers (\did{}s) provide generic decentralized identity without capability semantics~\cite{did-core}.
A \did{} identifies an entity but says nothing about what that entity can do.
Our capability path fills this gap.
An agent \uri{} could be viewed as a specialized \did{} with built-in capability description and discovery mechanism.
Where \did{}s are general-purpose, we are agent-specific---and that specificity enables capability-based discovery that generic \did{}s cannot support.

\subsection{Agent Discovery and Directory Services}

The Contract Net Protocol~\cite{smith1980contract} enables capability-based task allocation through manager-contractor negotiation.
This protocol assumes agents can already communicate---it does not address how the manager discovers potential contractors in the first place.
AgentNet decentralizes coordination among LLM agents through evolving directed-graph topologies~\cite{agentnet} and makes the same assumption: participants can already identify and reach one another.
Our scheme provides the discovery layer that both require.

\dns{}-SD and mDNS~\cite{dns-sd,rfc7558} work for local networks but do not scale to internet-wide agent discovery.
Our \dht{}-based approach provides comparable semantics (service type as capability path) at global scale.

Service mesh discovery systems like Consul~\cite{consul} provide infrastructure-level discovery with health checking.
These systems discover services, not agents with capabilities.
Our scheme adds semantic capability matching that infrastructure discovery cannot provide.

Closest to our work, the \nanda{} Index proposes internet-scale agent discovery through a federated ``quilt'' of registries~\cite{nanda-index}.
A lean, signed index record of at most 120 bytes maps an agent identifier to AgentFacts: self-describing verifiable-credential documents carrying capabilities, endpoints, and routing metadata.
\nanda{} specifies the registry infrastructure that our specification deliberately leaves open, including federation across administrative domains, sub-second revocation, endpoint agility, and privacy-preserving resolution.
What it leaves unspecified is naming.
Agent names in the index are ad-hoc handles (\texttt{@company:shop}, \texttt{urn:agent:...}) with no published grammar, normalization rules, or capability semantics, and the index tier cannot answer capability-scoped queries because capabilities live a tier below, in AgentFacts documents.
The designs are complementary: the \texttt{agent://} scheme is a candidate naming and identity layer for a \nanda{}-style index, contributing canonical names that carry trust and capability semantics plus deterministic index keys.
A recent survey situates registry approaches (centralized, enterprise, and distributed) in the broader landscape~\cite{agent-registry-evolution}.
DIAP explores an adjacent point in the design space: decentralized agent identity with zero-knowledge proofs over a hybrid peer-to-peer stack~\cite{diap}.

\subsection{Decentralized Systems}

Under Kademlia's standard assumptions, we inherit expected $O(\log N)$ routing for each exact-key lookup~\cite{kademlia}.
Our contribution is the key derivation scheme that maps trust-scoped capability paths to \dht{} keys, enabling semantic discovery over a content-addressed substrate.

Hewitt's actor model provides location transparency---actors communicate by reference without knowing physical location~\cite{hewitt1973actors,agha1986actors}.
Our scheme brings this property to internet-scale agent systems.
The agent \uri{} is an actor reference; resolution finds current location.

The libp2p networking stack provides a production Kademlia implementation and peer discovery~\cite{libp2p}.
Our reference implementation defines \dht{} operations against a transport-agnostic interface designed to admit a libp2p Kademlia backend; the discovery evaluation in Section~6 runs against an in-memory implementation of that interface.

\subsection{Trust and Attestation}

The W3C Verifiable Credentials data model provides general-purpose attestation~\cite{vc-data-model}.
Our \paseto{} tokens are a specialized form: the issuer is the trust root, the subject is the agent \uri{}, and the claims include capability paths.
We adopt VC's conceptual model with agent-specific semantics.
\nanda{}'s AgentFacts apply the VC model to agent metadata at internet scale~\cite{nanda-index}, and a recent IETF draft extends the Entity Attestation Token with capability claims for agentic systems~\cite{ietf-eat-cap}.
Both signal convergence on cryptographically verifiable capability claims; \paseto{} attestation is our deliberately minimal realization of the same requirement.
Zero-trust identity frameworks for agentic AI~\cite{zero-trust-agents} and decentralized trust fabrics for cross-domain credential exchange~\cite{trust-fabric} argue that agent trust must be continuously verified rather than assumed; the verification flow of Section~5.5 takes the same posture.

Dennis and Van Horn's capability model treats capabilities as unforgeable tokens granting specific permissions~\cite{dennis1966capabilities,miller2003capability}.
Our capability paths are descriptive (what the agent can do) rather than authoritative (permission to do it).
The attestation token is the authoritative element, binding the descriptive path to organizational endorsement.

Sabater and Sierra survey computational trust and reputation models~\cite{sabater2005trust}.
Our trust model is institutional rather than behavioral: trust roots are organizations making formal attestations, not agents accumulating reputation scores.
This matches enterprise deployment patterns where organizational accountability matters~\cite{ramchurn2004trust}.
Organizational models of multi-agent systems formalize institutions, roles, and norms~\cite{dignum2004agent,esteva2002islander}; the trust root serves as a lightweight institutional anchor in that tradition.

\subsection{Summary Comparison}

Table~\ref{tab:related-comparison} summarizes how existing approaches address the requirements from Section~2.3.

\begin{table}[h]
\centering
\small
\begin{tabular}{@{}lccccc@{}}
\toprule
\textbf{Approach} & \textbf{R1} & \textbf{R2} & \textbf{R3} & \textbf{R4} & \textbf{R5} \\
 & Topology & Capability & Decentral. & Scoping & Verifiable \\
\midrule
\fipa{} \df{} & No & Yes & No & No & No \\
A2A Protocol & No & No & No & No & No \\
\did{}s & Yes & No & Yes & No & Partial \\
\dns{}-SD & No & Yes & No & No & No \\
Consul & No & Partial & No & No & No \\
\nanda{} Index & Yes & Partial & Yes & Partial & Yes \\
\textbf{agent:// (ours)} & \textbf{Yes} & \textbf{Yes} & \textbf{Yes} & \textbf{Yes} & \textbf{Yes} \\
\bottomrule
\end{tabular}
\caption{Requirement satisfaction by approach. R1: topology independence; R2: capability-based discovery; R3: decentralized resolution; R4: organizational scoping; R5: cryptographically verifiable claims.}
\label{tab:related-comparison}
\end{table}

No existing approach satisfies all requirements.
\fipa{} \df{} provides capability semantics but centralizes control.
A2A addresses interoperability but binds identity to \uri{} endpoints.
\did{}s provide decentralized identity but lack capability discovery.
\dns{}-SD and Consul are infrastructure-level, not agent-level.
The \nanda{} Index comes closest: identity is topology-independent and claims are verifiable, but capability semantics live below the index tier and organizational scoping is registry policy rather than a structural property of the name.
The \texttt{agent://} scheme addresses all five requirements in one integrated design.


\section{Discussion and Limitations}

\subsection{Deployment Walkthrough}

The invoice-approval agent introduced in Section~1 provides a concrete example of the \texttt{agent://} scheme operating through its complete lifecycle.

\textbf{1. Creation.}
Acme Corp provisions an invoice approval agent, generating TypeID \texttt{agent\_01h455vb4pex5vsknk084sn02q}; \texttt{agent} is a generic extension class, while a deployment could instead select a core class such as \texttt{rule}.
The full \uri{} becomes \url{agent://acme.com/workflow/approval/invoice/agent_01h455vb4pex5vsknk084sn02q}.

\textbf{2. Attestation.}
Acme's trust root issues a \paseto{} token binding the agent to capability path \texttt{/workflow/approval/invoice} with 30-day expiration.
The attestation's \texttt{capabilities} claim includes this exact identity path; narrower descendant grants may be added for operations, but registration remains bound to the subject URI.

\textbf{3. Registration.}
The agent stores the same signed record at the keys for \texttt{workflow}, \texttt{workflow/approval}, and \texttt{workflow/approval/invoice}, together with its current endpoint (\texttt{https://agents.acme.com/v1/approvals}).

\textbf{4. Discovery.}
Partner organization Globex queries for \texttt{/workflow/approval} under \texttt{acme.com}.
The \dht{} returns registration records for all agents at that path and descendants, including Acme's invoice approval agent.

\textbf{5. Verification.}
Globex fetches Acme's public key from \texttt{https://acme.com/.well-known/agent-keys.json}, verifies the attestation signature, confirms the \texttt{capabilities} claim covers \texttt{/workflow/approval/invoice}, and checks expiration.
Verification succeeds; the agent is trusted.

\textbf{6. Interaction.}
Globex's procurement agent sends an invoice approval request to the discovered endpoint.
The agents communicate using A2A or other protocols; the \texttt{agent://} \uri{} provided stable identity for discovery.

\textbf{7. Migration.}
Six months later, Acme migrates to a new cloud provider.
The agent updates its \dht{} record with the new endpoint (\texttt{https://agents-new.acme.com/v1/approvals}).
The \uri{} remains unchanged.
Globex's cached reference still resolves---to the new endpoint.

\textbf{8. Attestation Refresh.}
At day 25, the agent requests a fresh attestation from Acme's trust root.
The old attestation remains valid until day 30; the new attestation is valid for the next 30 days.
No service interruption occurs.

This lifecycle demonstrates topology independence (step 7), capability-based discovery (step 4), and cryptographic verification (step 5) working together in a realistic enterprise scenario.

\subsection{When to Use Agent URIs}

The \texttt{agent://} \uri{} scheme is most valuable when:

\textbf{Migration is expected.}
If agents will move between cloud providers, scale across regions, or outlive their original infrastructure, stable identity prevents reference breakage.

\textbf{Cross-organization discovery is required.}
If agents from multiple trust roots must interact, \dht{}-based discovery with explicit trust scoping enables federation without centralized registries.

\textbf{Auditability matters.}
If audit logs must reference the same agent across time and infrastructure changes, stable \uri{}s provide consistent identification.

\textbf{Capability verification is needed.}
If agents' capability claims must be cryptographically verified rather than self-declared, the attestation model provides this assurance.

\subsection{Limitations}

\textbf{Reference implementation scope.}
The current backend is a single-process, in-memory simulation of the indexing contract, not a deployed Kademlia network. It materializes ancestor keys and enforces attestations, capacity, TTL, and immutable path binding, but it does not measure network routing, churn, replication, cache consistency, or well-known key discovery. A libp2p backend and end-to-end deployment evaluation remain future work.

\textbf{Corpus provenance.}
The artifact records source labels and the extracted definitions but not pinned upstream package versions or commits for every framework snapshot. The 369-tool result is reproducible from the bundled corpus, but independently reconstructing the same upstream snapshot is not yet guaranteed. Future releases should record package versions, extraction dates, and source commits.

\textbf{Trust-root discovery.}
The scheme deliberately requires callers to select one or more trust roots; it does not specify a global authority-discovery mechanism. Cross-root search therefore depends on application policy, federation configuration, or a higher-level directory.

\textbf{Trust root infrastructure required.}
Organizations must operate trust root infrastructure: key management, attestation issuance, well-known endpoint hosting.
This is comparable to operating a certificate authority for TLS, but adds operational burden.

\textbf{\dht{} participation incentives.}
The scheme relies on \dht{} nodes storing and serving registration records.
Why would operators run \dht{} nodes?

Three incentive models apply:

\begin{enumerate}
\item \textbf{Reciprocity-based.} Nodes only serve queries from peers who also serve them (tit-for-tat).
BitTorrent's Mainline \dht{} demonstrates this model's viability at scale: millions of nodes participate because reciprocity is enforced at the protocol level.

\item \textbf{Stake-based.} Trust roots stake reputation or deposit to participate; misbehavior (dropping queries, serving false records) forfeits stake.
This requires a coordination layer beyond the base \dht{} protocol.

\item \textbf{Utility-derived.} Organizations participate because discovery of their own agents requires \dht{} presence.
If you want others to find your agents, you must contribute to the network that enables discovery.
IPFS demonstrates this model: organizations running IPFS gateways gain discoverability for their content.
\end{enumerate}

The specification does not mandate an incentive model---this is a deployment-time policy choice.
For enterprise deployments, utility-derived incentives may suffice (organizations have business reasons to be discoverable).
For open deployments, reciprocity or stake mechanisms may be needed to prevent free-riding.
This is an underspecified aspect of the design, noted as future work.

\textbf{No global capability ontology.}
Each trust root defines its own capability namespace.
If \texttt{acme.com/workflow/approval} and \texttt{globex.com/process/authorize} denote the same capability, cross-organization discovery fails silently---queries find only exact path matches.

This is intentional: avoiding global governance prevents political capture and governance bottlenecks.
However, it limits plug-and-play federation.

A lightweight coordination mechanism could address this: \emph{Capability Mapping Services}.
Third parties (industry consortia, standards bodies) publish equivalence mappings:

\begin{lstlisting}
{
  "acme.com/workflow/approval": [
    "globex.com/process/authorize",
    "initech.com/approvals/workflow"
  ]
}
\end{lstlisting}

Requesters consult mapping services to expand queries across equivalent paths.
No global ontology is required; mappings are opt-in and domain-specific.
This approach has advantages over a global ontology: domain expertise stays with domain experts, there is no governance bottleneck, and the system degrades gracefully (works without mappings, better with them).

\textbf{Mapping Service Discovery.}
How does a requester discover relevant mapping services?
We propose a layered approach:
(1)~Trust roots optionally publish their own mappings at a well-known endpoint: \texttt{https://\{trust-root\}/.well-known/capability-mappings.json}.
(2)~Trust roots may reference third-party mapping services they endorse.
(3)~Industry registries (analogous to CA certificate lists) can aggregate mapping services for specific domains---healthcare, finance, logistics---providing curated equivalences for their communities.

\textbf{Conflict Resolution.}
When multiple mapping services provide conflicting information (mapping service A says path X equals Y, mapping service B says X does \emph{not} equal Y), resolution follows requester policy.
If mappings agree, union semantics apply: query all equivalent paths and merge results.
If mappings contradict, the requester decides which mapping service to trust for their use case.
This is a feature, not a bug: different communities may legitimately have different equivalence judgments, and no global authority should override local domain expertise.

Standardizing capability mapping service discovery and format is noted as future work.

\textbf{\dht{} poisoning risk.}
Malicious actors could register fake agents at popular capability paths.
Attestation verification filters invalid registrations, but verification requires fetching keys from potentially malicious trust roots.
Rate limiting and proof-of-work could mitigate but are not specified.

\subsection{Future Work}
\label{sec:future}

\textbf{Privacy-preserving discovery.}
Can agents register capabilities without revealing identity until matched?
Commitment schemes and onion routing could enable this, but tension with attestation verification remains.

\textbf{Capability ontology standardization.}
A shared vocabulary for common capabilities would improve cross-organization interoperability.
Who maintains such an ontology and how it evolves are open governance questions.

\textbf{Integration with A2A.}
The A2A protocol is gaining adoption~\cite{a2a-spec}.
A2A Agent Cards contain an \texttt{endpoints} field with service \uri{}s for communication.
We propose extending Agent Cards with an \texttt{agent\_uri} field providing stable identity:

\begin{lstlisting}
{
  "name": "Invoice Approval Agent",
  "agent_uri": "agent://acme.com/workflow/approval/agent_01h...",
  "endpoints": [
    {"url": "https://agents.acme.com/v1/approval", "protocol": "https"},
    {"url": "grpc://agents.acme.com:50051", "protocol": "grpc"}
  ],
  "attestation": "v4.public.eyJpc3MiOiJhY21lLmNvbSIs..."
}
\end{lstlisting}

The integration flow:
(1)~Requester discovers agents via \texttt{agent://} \uri{} resolution, obtaining \dht{} records.
(2)~Records include Agent Card \uri{} or embedded Agent Card.
(3)~Agent Card provides current endpoints (HTTPS, gRPC, WebSocket) for communication.
(4)~Attestation token in the Agent Card enables capability verification.

A2A provides communication semantics (how to talk to agents); the \texttt{agent://} scheme provides identity semantics (which agent you're talking to).
The combination yields stable identity with flexible transport---agents can update endpoints without breaking identity references.

Specifying this integration formally and building reference implementations would accelerate practical deployment.


\section{Conclusion}

Agent systems today suffer from a fundamental architectural flaw: identity is bound to location.
When agents migrate, references break.
When organizations federate, discovery requires centralized coordination.
When capabilities must be verified, self-declaration is the only option.

The \texttt{agent://} \uri{} scheme addresses these problems through principled separation of concerns.
The trust root establishes organizational authority.
The capability path describes what the agent does and is constitutive identity material; changing it creates a new agent identity.
The agent identifier provides stable, sortable reference.
Together, these components create topology-independent identity with built-in capability semantics.

\dht{}-based discovery enables finding agents by what they do, not where they are.
Trust-root scoping prevents cross-organization pollution while permitting explicit federation.
\paseto{} attestation binds capability claims to cryptographic verification, moving beyond self-declaration to institutional accountability.

Our evaluation demonstrates that these properties hold in practice.
The capability grammar achieves 100\% coverage on real-world tools.
The simulated ancestor-key index exactly reproduces generated lookup ground truth across 10,000 registrations.
Endpoint-migration identity stability holds by construction.
Measured local parsing and key-derivation primitives complete in microseconds.

The \texttt{agent://} \uri{} scheme provides a formally-specified, practically-evaluated foundation for decentralized agent identity.
Like \dns{} separated network names from addresses, enabling the internet we know today, we believe capability-addressed agent identity can enable the multi-agent systems of tomorrow.

\bibliographystyle{ACM-Reference-Format}
\bibliography{references}

\end{document}